\documentclass[10pt,letterpaper]{article}
\usepackage{amsmath}
\usepackage{amssymb}
\usepackage{cite}

\usepackage{opex3}
\usepackage{txfonts}
\usepackage{amssymb}
\usepackage{tipa}
\usepackage{graphicx}
\usepackage{txfonts}
\usepackage{amssymb}
\usepackage{extarrows}

\usepackage{color}
\definecolor{MyDarkBlue}{rgb}{0,0.08,0.45}
\definecolor{yellow}{rgb}{0.99,0.99,0.70}
\definecolor{white}{rgb}{1.0,1.0,1.0}
\definecolor{black}{rgb}{0.00,0.00,0.00}
\definecolor{green}{rgb}{0.8,0.98,0.83}
\pagecolor{green}
\begin{document}
\title{Shortcuts to adiabaticity in non-Hermitian quantum systems without rotating-wave approximation}

\author{Hong Li,$^{1}$ H. Z. Shen,$^{1,2}$ S. L. Wu $^{3}$ and X. X. Yi*$^{1,2,}$}
\address{$^1$Center for Quantum Sciences and School of Physics,
Northeast Normal University, Changchun 130024, China\\
$^2$Center for Advanced Optoelectronic Functional Materials
Research, and Key Laboratory for UV Light-Emitting Materials and
Technology of Ministry of Education, Northeast Normal University,
Changchun 130024, China\\
$^3$School of Physics and Materials Engineering, Dalian Nationalities University, Dalian 116600, China}

\email{$^{*}$yixx@nenu.edu.cn}

\begin{abstract}
The technique of shortcuts to adiabaticity (STA) has attracted broad attention due to their possible applications in quantum information processing and quantum control. However,
most studies published so far have  been only focused on Hermitian systems under the rotating-wave approximation (RWA). In this paper, we propose a modified shortcuts to adiabaticity
technique to realize population transfer for a non-Hermitian system without RWA. We work out an exact expression for the control function and present examples consisting of two- and three-level
systems with decay to show the theory. The results suggest that the shortcuts to adiabaticity technique presented here is robust for fast passages. We also find that the decay has small effect on
the population transfer in the three-level system. To shed more light on the physics behind this result, we reduce the quantum three-level system to an effective two-level one with large detunings.
The shortcuts to adiabaticity technique of effective two-level system is studied. Thereby the high-fidelity population transfer can be implemented in non-Hermitian systems by our method, and it works
even without RWA.
\end{abstract}

%%%%%%%%%%%%%%%%%%%%%%% References %%%%%%%%%%%%%%%%%%%%%%%%%

%%%%%%%%%%%%%%%%%%%%%%%%%%  body  %%%%%%%%%%%%%%%%%%%%%%%%%%
\section{Introduction}\label{s2}

Population transfer by adiabatic evolution is among the most popular methods for coherent atomic manipulation
~\cite{Vitanov4535,shiyan023821,Wei113601,Halfmann7068,Bergmann1003,LG033419,Sun033408,Wiebe013024,Takahashi315304,Torosov043418}.
However, due to the requirement on  adiabaticity, the manipulation is very slow. Recently, by designing nonadiabatic shortcuts to
speed up quantum adiabatic process, a new technique named``shortcuts to adiabaticity" (STA)
~\cite{chenxi033405,STA123003,STA9937,STA25707,STA1630,chen043402,Paul052303,xiayan052109,Campbell177206,Okuyama070401,STA052303,Kumar10628}
opens a new door towards to fast and robust quantum state manipulations. The shortcut techniques include counterdiabatic
control protocols, transitionless quantum driving(TQD)~\cite{TQD053408,TQD015025,TQD100403,TQD052109,Takahashi062117,
LFWei015204,LRchenxi062116,Songxueke052324,Berry2009},``fast-forward" scaling ~\cite{Masuda043434} , and inverse
engineering based on Lewis-Riesenfeld (LR) invariants ~\cite{Kiely115501,Masuda033621}. Among the methods of shortcuts to adiabaticity, transitionless quantum driving
has been intensively studied. The key idea of transitionless quantum driving is to use a shortcuts to adiabaticity by adding extra fields to cancel the nonadiabatic couplings.

The shortcuts to adiabaticity technique has proved to be successful in designing control scheme for closed quantum systems. In practice, however,
interactions between a quantum system and its surroundings is not avoidable, leading to coherence loss of the closed systems. Such a system can be
described by a non-Hermitian Hamiltonian, and in the past few decades, it has drawn much attention ~\cite{NHMoiseyev2011}, especially in the context
of PT-symmetric systems ~\cite{PTBender957,PTMostafazadeh7081}. There are many amazing phenomena predicted in PT-symmetric systems, for instance,
a PT-symmetric Hamiltonian possesses a higher quantum speed efficiency than Hermitian Hamiltonian systems ~\cite{Uzdin415304,Hang083604}. Recently,
an approximation of the adiabatic condition for non-Hermitian systems were derived ~\cite{Ibanez033403}. For non-Hermitian systems the usual approximations
and criteria are not valid in general, so results that are applicable for Hermitian systems have to be reconsidered and modified. The extension of shortcuts to adiabaticity
from Hermitian systems to non-Hermitian systems has been put forward for the Landau-Zener(LZ) model ~\cite{Reyes444027}. Therefore, we may wonder \emph{if there is a general shortcuts to adiabaticity
technique for non-Hermitian PT-Hamiltonian systems? whether the adiabatic evolution proposed in Ref.\cite{Ibanez033403} can be accelerated for such systems?}

On the other hand, the rotating wave approximation (RWA) is in common used in atomic optics and magnetic resonance ~\cite{Chan065507,Irish259901,Klimov063811},
where the counter-rotating (CR) terms are neglected. The RWA is valid in the weak coupling regime with small detuning and weak field amplitude, in this case the
contribution of the counter-rotating terms to the evolution of the system is quite small. In recent experiments, efforts are made to reach a
promising method for studying strong- and ultrastrong-coupling physics ~\cite{Liu54003,Ashhab042311,Casanova263603,Fedorov060503,Irish173601,Bourassa032109}, where the RWA is no longer valid.
Refs. ~\cite{Larson033601,Sun012107} found that the RWA may lead to faulty results. Especially, recent developments in physical implementation lead to strong coupling between qubit and cavity
modes ~\cite{You589,Niemczyk772}, which requires a careful consideration of the effect of counter-rotating terms. This motivate us to ask:
\emph{if can we design a shortcuts to adiabaticity technique for an open system with the counter-rotating terms?}

In this paper, we propose a renewed shortcuts to adiabaticity technique for non-Hermitian quantum systems beyond RWA. By using transitionless quantum driving method, we determine the exact
control to speed up the adiabatic population transfer. Then, we apply it into the two- and three-level systems with decay and without RWA. We numerically calculate the population transfer
dynamics with and without counter-rotating terms. The results show that the population transfer with the counter-rotating terms are more steady. On the other hand, we also find that the decay
of the excited state has small effect on the population transfer in the three-level system. This is attributed to the fact that the population transfer $| 1 \rangle \rightarrow | 3 \rangle$ is realized
by the dark state. In the case of large detuning, we reduce the three-level system to an effective two-level system. The shortcuts to adiabaticity technique is then applied to the
effective two-level system, which might shed more light on the present scheme.

This paper is organized as follows. In Sec.~\ref{two}, we study the shortcuts to adiabaticity applied to a decaying ~\cite{decay053406,decay095303,decay13727}
two-level system without RWA. We numerically compare the population transfer with and without counter-rotating terms. In Sec.~\ref{three}, we study the shortcuts
to adiabaticity applied to a decaying \cite{xjdecay043001,xjdecay053415} three-level system without RWA. We numerically calculate the population transfer with and without
counter-rotating terms. The effect of the decay on the population transfer in the three-level system is also discussed in this section. Sec.~\ref{four} is devoted to discussion and conclusion.

\section{Transitionless driving applied to a decaying two-level atom beyond the RWA} \label{two}

In this section, we present a shortcuts to adiabaticity technique for an open two-level quantum system beyond RWA. Assume the
two-level system interacts with a laser electric field with linear polarization in $x$ direction $\vec
E(t)=E_{0}(t)\cos[\omega_{L}t]\vec x$, the Hamiltonian of the
two-level system reads,
\begin{eqnarray}
H_{S}(t)&=&\frac{\hbar}{2} [\Omega_{R}(t)(|2\rangle\langle1|+|1\rangle\langle2|)(e^{i\omega_{L}t}+e^{-i\omega_{L}t})\\ \nonumber
&&+\omega_{0}(t)(|2\rangle\langle2|-|1\rangle\langle1|)],
\label{sss}
\end{eqnarray}
where $|1\rangle$ ($|2\rangle$ ) is the ground (excited) state of
the two-level atom, $\omega_{0}$ is the transition frequency,
$\Omega_{R}(t) $ and $\omega_{L}$ are the strength and the
frequency of the classical field. Defining
\begin{eqnarray}
H_{L}(t)&=& i \hbar \dot{U}(t)U^{\dagger}(t)\\ \nonumber &=& \frac{\hbar \omega_{L}}{2} (|2\rangle\langle2|-|1\rangle\langle1|),
\label{xg2}
\end{eqnarray}
with unitary operator
\begin{eqnarray}
U(t)&=&e^{-\frac{i}{\hbar}\int_{0}^{t}H_{L}(t')dt'}\\ \nonumber &=&e^{-i\omega_{L}t/2}|2\rangle\langle2|-e^{i\omega_{L}t/2}|1\rangle\langle1|,
\label{xg3}
\end{eqnarray}
we can transform the Hamiltonian (1) into the interaction picture, i.e.,
\begin{eqnarray}
H(t) &=&U^{\dagger}(t)[H_{S}(t)-H_{L}(t)]U(t)\nonumber\\&= &\frac{\hbar }{2} \left [ {\begin{array}{*{20}{c}}
-\Delta (t)&{\Omega (t)}\\
{\Omega (t)}&{\Delta (t) }
\end{array}} \right].
\label{xg4}
\end{eqnarray}
Here the bases are  $|1\rangle  = {[1,0]^T}$ and $|2\rangle
={[0,1]^T}$ with $T$ denoting transposition.\\
With
\begin{eqnarray}
\Omega (t)=\Omega_{R} (t)(1 + {e^{-2i \omega_{L} t}}),
\label{xg5}
\end{eqnarray}
and
\begin{eqnarray}
\Delta(t)=\omega_{0}(t)-\omega_{L},
\label{xg6}
\end{eqnarray}
in which ${e^{ \pm 2i{\omega _L}t}}$ are the counter-rotating terms. To describe the decay of the upper level, we add an
imaginary energy  $i\Gamma$ to the energy of the excited state. The Hamiltonian then becomes non-Hermitian,
\begin{eqnarray}
{{\tilde H}_{0}}(t) = \frac{\hbar }{2} \left [ {\begin{array}{*{20}{c}}
-\Delta (t)&{\Omega (t)}\\
{\Omega (t)}&{\Delta (t)-i\Gamma }
\end{array}} \right].
\label{xg7}
\end{eqnarray}
The dynamics of the two-level quantum system is described by the Schr\"odinger equation,
\begin{eqnarray}
i\hbar \frac{\partial }{{\partial t}} \left[{\begin{array}{*{20}{c}}
{{C_1}(t)}\\
{{C_2}(t)}
\end{array}} \right] = {{\tilde H}_{0}}(t)  \left [{\begin{array}{*{20}{c}}
{{C_1}(t)}\\
{{C_2}(t)}
\end{array}} \right],
\label{relhh}
\end{eqnarray}
where ${{C_1}(t)}$ and ${{C_2}(t)}$ are probability amplitudes of the two bare states $| 1 \rangle $ and $| 2 \rangle $, respectively.

To design a shortcut to adiabaticity for this non-Hermitian system without RWA, we have to explore the left and right eigenstates of
the non-Hermitian Hamiltonian (\ref{xg7}). Straightforward calculations lead to left eigenstates of $\tilde H_0(t)$ as
\begin{eqnarray}
\begin{aligned}
| \varphi_{+}(t) \rangle  &=\sin \beta| 1 \rangle  + \cos \beta e^{i\omega_{L}t}| 2 \rangle,&\\
| \varphi_{-}(t) \rangle  &=\cos \beta e^{-i\omega_{L}t}| 1 \rangle  -\sin \beta| 2 \rangle,&
\end{aligned}
\label{re}
\end{eqnarray}
where the mixing angle $\beta=\beta (t)$ is complex and defined as
\begin{eqnarray}
\beta=\frac{1}{2}\arctan\frac{2\Omega_{R}\cos\omega_{L}t}{\Delta-i\Gamma/2},
\label{re11}
\end{eqnarray}
and the right eigenstates can also be determined,
\begin{eqnarray}
\begin{aligned}
| \hat{\varphi}_{+}(t) \rangle  &=\sin \beta^{\ast}| 1 \rangle  + \cos \beta^{\ast} e^{i\omega_{L}t}| 2 \rangle,&\\
| \hat{\varphi}_{-}(t) \rangle  &=\cos \beta^{\ast}
e^{-i\omega_{L}t}| 1 \rangle  -\sin \beta^{\ast} | 2 \rangle,&
\end{aligned}
\label{re12}
\end{eqnarray}
where the asterisk means ¡°complex conjugate¡±.

By the spirit of the shortcut to adiabaticity ~\cite{TQD053408,TQD015025,TQD100403,TQD052109,Takahashi062117, LFWei015204,LRchenxi062116,Songxueke052324,Berry2009}, the
counterdiabatic term can be given by ~\cite{chenxi023415}
\begin{equation}
\begin{aligned}
{\tilde{H}_{CD}}(t) =& i\hbar [| {{\partial _t}\varphi_{+}(t)}
\rangle \langle {\hat{\varphi}_{+}(t)} |-\langle
{\hat{\varphi}_{+}(t)} | {{\partial _t}\varphi_{+}(t)} \rangle |
\varphi_{+}(t) \rangle \langle {\hat{\varphi}_{+}(t)} |
\\  & +| {{\partial _t}\varphi_{-}(t)} \rangle \langle
{\hat{\varphi}_{-}(t)} |-\langle {\hat{\varphi}_{-}(t)} |
{{\partial _t}\varphi_{-}(t)} \rangle | \varphi_{-}(t) \rangle
\langle {\hat{\varphi}_{-}(t)} |], \label{reh13}
\end{aligned}
\end{equation}
where
\begin{eqnarray}
\begin{aligned}
&\langle {\hat{\varphi}_{\pm}(t)} |  {{\partial _t}\varphi_{\pm}(t)} \rangle = \pm i\omega_{L} \cos^{2}\frac{\beta}{2},&\\
&\langle {\hat{\varphi}_{\mp}(t)} |  {{\partial _t}\varphi_{\pm}(t)} \rangle = (\pm \frac{\dot{\beta}}{2}-i\frac{\omega_{L}}{2}\sin\beta)e^{\pm i \omega_{L} t},&
\end{aligned}
\label{re14}
\end{eqnarray}
which is equivalent to
\begin{equation}
\begin{aligned}
{\tilde{H}_{CD}}(t) =\hbar \left [ {\begin{array}{*{20}{c}}
{\frac{\omega_L}{2}\sin^{2}}\beta &{(i\frac{\dot{\beta}}{2}+\frac{\omega_{L}}{4}\sin2\beta)e^{-i\omega_{L}t}}\\
{(-i\frac{\dot{\beta}}{2}+\frac{\omega_{L}}{4}\sin2\beta)e^{i\omega_{L}t}}&{-\frac{\omega_L}{2}\sin^{2}\beta}
\end{array}}\right ].
\label{reh15}
\end{aligned}
\end{equation}
Here, we should stress that the counterdiabatic terms presented in Eq.~(\ref{reh15}) is beyond RWA. However, we can easily obtain the
counterdiabatic term for the RWA case by ignoring the counter-rotating terms in $H_0(t)$. Applying the rotating wave
approximation to get rid of the counter-rotating terms, we end up with
\begin{eqnarray}
{H_0}(t) = \frac{\hbar }{2} \left [ {\begin{array}{*{20}{c}}
-\Delta (t)&{\Omega_{R} (t)}\\
{\Omega_{R} (t)}&{\Delta (t)-i\Gamma }
\end{array}} \right].
\label{xg8}
\end{eqnarray}
%%%%%%%%%%%%%%%%%%%%%%%%%%%%%%%%%%%%%%%%%%%%%%%%%%%%%%%%%%%%%%%%%%%%%%%%%%%%%%%%%%%%%%%%%%%%%%%%%%%%%%%%%%%%%%%%%%%%%%%%%%%%%%%%%%%%%%%%%%%%%%%%%%%%%%%%%%%%%%%%%%%%%%%%%%%%%%%%%%%%%%%%%%%%%%%%%%
%figure 1
\begin{figure*}[tbp]
\centering
\includegraphics[width=0.45\textwidth]{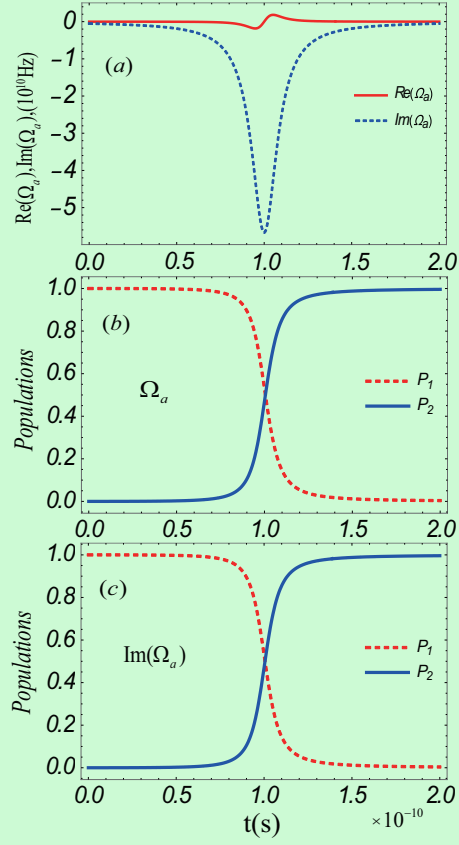}
\caption{Real (red solid line) and imaginary (blue dashed line) parts of ${{\Omega _a}}$ (a). The population evolution is driven by  ${{\Omega _a}}$ (b), and [$Im{{\Omega _a}}$]
(c). The parameters chosen are $\Gamma = 2 \pi \times 0.5 Mhz$, ${\Omega _0} = 2 \pi \times 5 Mhz$, $\delta  = 2 \pi \times 300 Mhz$, ${t_f}=2 \times 10^{-10}s$, ${t_0}=10^{-10}s$.}
\end{figure*}
%%%%%%%%%%%%%%%%%%%%%%%%%%%%%%%%%%%%%%%%%%%%%%%%%%%%%%%%%%%%%%%%%%%%%%%%%%%%%%%%%%%%%%%%%%%%%%%%%%%%%%%%%%%%%
%figure 2
\begin{figure*}[tbp]
\centering
\includegraphics[angle=0,width=0.40\textwidth]{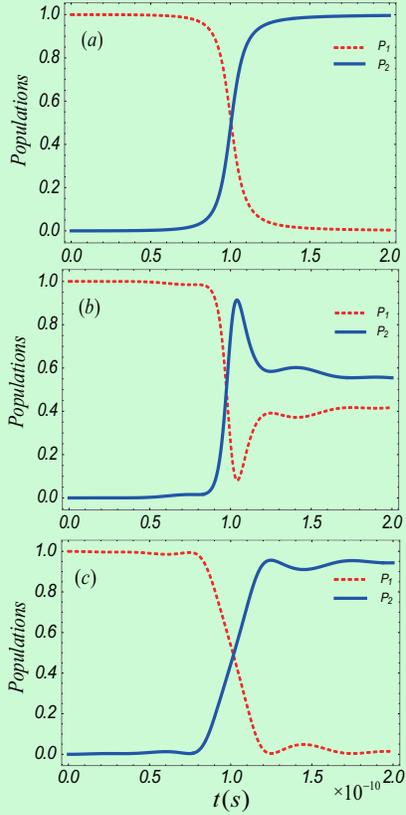}
\caption{The evolution of the population for a ndecaying two-level system with decay rate $\Gamma = 2 \pi \times 0.5 MHz$. The population transfer is implemented by the shortcuts to adiabaticity
technique under RWA [Figs. 2(a), and 2(b)], and implemented by the shortcuts to adiabaticity technique beyond RWA [Fig. 2(c)]. Note that all the counter-rotating terms are neglected in Fig. 2(a),
but are considered in Fig. 2(b). The population of level $|1 \rangle $ (dashed red), and $|2 \rangle $ (solid blue) are shown. The other parameters: $\Omega _0 = 2 \pi \times 5 MHz$,
$\delta = 2 \pi \times 300 MHz$, $ t_f = 2 \times 10^{-10}s $, $t_0 = 10^{-10}s $, $\omega _{L} = 2 \pi \times 10 GHz $.} \label{fig2}
\end{figure*}
By the same procedure as above, $H_{CD}$ reads
\begin{eqnarray}
{H_{CD}}(t) = {\hbar }\left [ {\begin{array}{*{20}{c}}
0&{\Omega_a (t)}\\
{ - \Omega_a (t)}&0
\end{array}}\right ],
\label{reh1nrwa}
\end{eqnarray}
where $\Omega_a (t)=i \dot{\theta} /2$ is the auxiliary pulse with
\begin{eqnarray}
\dot{\theta}=\frac{\dot{\Omega}_{R}[\Delta (t)-i \Gamma /2]-\Omega_{R}(t)(\dot{\Delta}-i\dot{\Gamma}/2)}{\Omega_{R}^{2}(t)+[\Delta (t)-i \Gamma /2]^{2}},
\label{reh1n}
\end{eqnarray}
which plays the role of the Rabi frequency for a fast-driving field. In principle, ${H_{CD}}(t)$ can drives the dynamics along the instantaneous eigenstates of ${H_0}(t)$ in arbitrarily
short time, but due to the practical limitations, e.g., the laser power, the performance would depend on those limitations.

The Hamiltonian ~(\ref{reh1nrwa}) has off-diagonal terms with real and imaginary parts, see Fig. 1. Practically, realizing the Hamiltonian ~(\ref{reh1nrwa}) is not straightforward. From the
Hamiltonian ~(\ref{reh1nrwa}), we can see that the off-diagonal terms are not complex conjugate to each other except that the real part of $ {{\Omega _a}}$ is zero. The imaginary parts, shown by
the dip in Fig. 1(a), can be  realized  by a complementary laser with orthogonal polarization ~\cite{STA123003}, the real parts constitute a non-Hermitian contribution to the Hamiltonian. They
are, however, contribute very small to the dynamics. Ignoring the real part of $\Omega_{a}$, i.e., setting $\Omega_{a} \simeq i Im[\Omega_{a}]$, we plot the Figs. 1(b) and 1(c). We can see that
the population transfer is almost the same. So the Hamiltonian ~(\ref{reh1nrwa}) can be realized approximately. When the real parts can not be ignored, the realization mentioned above is not valide anymore.

In the RWA regime, i.e., drop the terms with $e^{\pm 2i\omega_{L}t}=0$. It can be verified that Eq.~(\ref{reh15}) is the same as Eq.~(\ref{reh1nrwa}). Figure ~\ref{fig2} shows the population of the
ground state $|1\rangle$ (red dash lines) and the excited state $|2\rangle$ (blue solid lines) with and without RWA. In our case, we use the shortcuts to adiabaticity technology to speed up the passage,
in which pulses do not need satisfy adiabaticity condition. It requires an additional laser field. Therefore we provide an example in which the adiabaticity condition fails and then apply
the Hamiltonian in Eq.~(\ref{reh1nrwa}) to remedy this problem and achieve a fast full decay. Here, the coherent controls in the Hamiltonians $\tilde H_0(t)$ and $H_0(t)$ are chosen to be the Allen-Eberly
(AE) drivings ~\cite{chenxi062163}, i.e.,
\begin{eqnarray}
\begin{aligned}
&\Omega_{R} (t) = {\Omega _0}{\mathop{\rm sech}\nolimits} \left [\frac{\pi (t-t_{f}/2)}{2{t_0}}\right ],&\label{omegar}\\
&\Delta (t) = \frac{{2{\delta ^2}{t_0}}}{\pi }\tanh \left [\frac{\pi (t-t_{f}/2)}{2{t_0}}\right ].&
\end{aligned}
\label{lhn}
\end{eqnarray}
Here, we select the appropriate parameters ~\cite{Ibanez033403,chenxi023415}, in the RWA regime, we use Hamiltonian $ H_0(t)$ and the counterdiabatic Hamiltonian $H_{CD}(t)$ to describe the two-level system
(see Fig. ~\ref{fig2}(a)). We find that the population is transferred from the ground state into the excited state completely. However, when the RWA fails, counter-rotating terms must be taken into account,
$H_{CD}(t)$ can not be used as the counterdiabatic Hamiltonian to accelerate the adiabatic dynamics governed by the reference Hamiltonian $\tilde H_0(t)$, see Fig.~\ref{fig2}(b). Therefore, in the regime beyond
RWA, we have to redesign the counterdiabatic terms of the shortcuts to adiabaticity according to Eq.~(\ref{reh15}). As shown in Fig.~\ref{fig2}(c), if we apply Hamiltonian (\ref{reh15}) into the dynamics governed by Hamiltonian $\tilde H_0(t)$, the population transfer efficiency can be enhanced remarkably. Our results suggest that shortcuts to adiabaticity technique well developed for Hermitian systems can be
extended to non-Hermitian one. It is envisaged that our study could stimulate further studies of coherent population transfer techniques of multilevel systems for non-Hermitian systems. In the next section, we mainly
discuss three-level non-Hermitian system.
%%%%%%%%%%%%%%%%%%%%%%%%%%%%%%%%%%%%%%%%%%%%%%%%%%%%%%%%%%%%%%%%%%%%%%%%%%%%%%%%%%%%%%%%%%%%%%%%%%%%%%%%%%%%%%%%%%%%%%%%%%%%%%%%%%%%%%%%%%%%%%%%%%%%%%%%%%%%%%%%%%%%%%%%%%%%%%%%%%%%%%%%%%%%%%%%%%%%%%%%%
\section{Transitionless driving scheme applied to a decaying three-level atom beyond RWA} \label{three}
\subsection{$H_{CD}(t)$ applied to a decaying three-level atom}

In this section, we consider a three-level system of $\Lambda$-type with decay in the excited state. The two lower states are long-lived. We consider the state $| 1 \rangle$ is initially populated, and the other
state $| 3 \rangle$ as the target state, and $|2\rangle $ as the excited state. Levels $|1\rangle$ and $|2\rangle$ are coupled by the Pump laser $\bar{\Omega}_{p}(t)$, while Levels $|2\rangle$ and $|3\rangle$
are coupled by the Stokes laser  $\bar{\Omega}_{s}(t)$. We assume that there is a decay in the excited state $|2\rangle $ with rate $\Gamma$. Such a system can be described by the following time-dependent Schr\"odinger equation,
\begin{eqnarray}
i\hbar \frac{d}{{dt}}c(t) = \bar{H}_0(t)c(t),
\label{lhx}
\end{eqnarray}
where the vector $c(t) = {[ {c_{1}(t),c_{2}(t),c_{3}(t)}]^T}$ is constructed by the three probability amplitudes of states $| 1 \rangle $, $| 2 \rangle $, and $| 3 \rangle $.
Without the rotating wave approximation, the reference Hamiltonian of the $\Lambda$-type three-level system can be depicted as ~\cite{xjdecay043001,xjdecay053415,xjdecay015006}
\begin{eqnarray}
\bar{H}_{0}(t) = \frac{\hbar }{2}\left [{\begin{array}{*{20}{c}}
0&\bar{\Omega}_{p}(t)&0\\
\bar{\Omega}^{\ast}_{p}(t)&{2{\Delta _p} - i\Gamma }&\bar{\Omega}_{s}(t)\\
0&\bar{\Omega}^{\ast}_{s}(t)&{2({\Delta _p} - {\Delta _s})}
\end{array}}\right ],
\label{lhy}
\end{eqnarray}
where the detunings are defined by ${\Delta _p} = {\omega _p} - {{({E_2} - {E_1})} \mathord{/ {\vphantom {{({E_2} - {E_1})} \hbar }} \kern-\nulldelimiterspace} \hbar }$, ${\Delta _s} = {\omega _s}
- {{({{E_2} - {E_3}})} \mathord{/ {\vphantom {{({{E_2} - {E_3}})} \hbar }} \kern-\nulldelimiterspace} \hbar }$ for the $\Lambda $ configuration. $\bar{\Omega}_{p}(t)=\Omega_{p}(t)(1+e^{-2i\omega_{p}t})$ and
$\bar{\Omega}_{s}(t)=\Omega_{s}(t)(1+e^{-2i\omega_{s}t})$ are the pump and Stokes pulses (beyond RWA). The imaginary term $ \Gamma$ describes the decay from $| 2 \rangle$. Conventional stimulated
Raman adiabatic passage (STIRAP) relies on the two-photon resonance, i.e., $\Delta=\Delta_p=\Delta_s$.
%%%%%%%%%%%%%%%%%%%%%%%%%%%%%%%%%%%%%%%%%%%%%%%%%%%%%%%%%%%%%%%%%%%%%%%%%%%%%%%%%%%%%%%%%%%%%%%%%%%%%%%%%%%%%%%%%%%%%%%%%
%figure 3
\begin{figure*}[tbp]
\centering
\includegraphics[angle=0,width=0.45\textwidth]{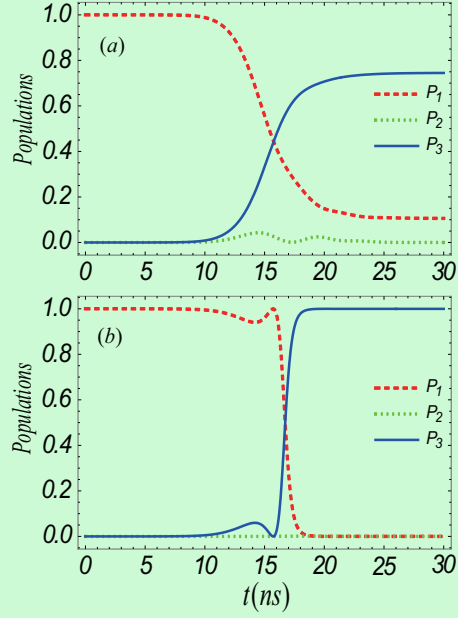}
\caption{Transition probability ${P_{1 \to 3}}$ of the fast stimulated Raman by the shortcuts to adiabaticity under and beyond RWA. Fig.~3(a) is plotted by using the shortcuts to
adiabaticity technique under RWA, but consider the CR terms. Fig.~3(b) is plotted by using the shortcuts to adiabaticity technique beyond RWA. We set the parameters: $t_{f}=30 ns$,
$T={t_f}/6$, $\tau={t_f}/10$, $\Omega_{0}=2 \pi \times 200 MHz$, $\Delta=2 \pi \times 200 MHz$, $\Gamma=2 \pi \times 100 MHz$, $\omega_{p}=100MHz$, $\omega_{s}=80MHz$. The dashed red,
dotted green, and solid blue lines describe the time-dependent population of the state $|1 \rangle, |2 \rangle$, and $|3 \rangle$, respectively.}
\label{fig3}
\end{figure*}

The mechanism of the population transfer in STIRAP and the limitation of adiabaticity can be easily understood by introducing the so-called adiabatic states. They are connected to the bare
states $\left| 1 \right\rangle $, $\left| 2 \right\rangle $ and $\left| 3 \right\rangle $ by the eigenstates of $\bar H_0(t)$,
\begin{eqnarray}
\begin{aligned}
| {{E_0}(t)} \rangle  &= \frac{1}{{{\Xi _0}(t)}}[\bar{\Omega}_{s}(t)| 1 \rangle  - \bar{\Omega}^{\ast}_{p}(t)| 3 \rangle ],& \\
| {{E_ + }(t)} \rangle &= \frac{1}{{{\Xi _1}(t)}}[\bar{\Omega}_{p}(t)| 1 \rangle  + {\varepsilon _ + }(t)| 2 \rangle  + \bar{\Omega}^{\ast}_{s}(t)| 3 \rangle ],&\\
| {{E_ - }(t)} \rangle &= \frac{1}{{{\Xi _2}(t)}}[\bar{\Omega}_{p}(t)| 1 \rangle  + {\varepsilon _ - }(t)| 2 \rangle  + \bar{\Omega}^{\ast}_{s}(t)| 3 \rangle ],&
\end{aligned}
\label{lhz}
\end{eqnarray}
where ${\Xi _0}(t) = \sqrt {{{| \bar{\Omega}_{p}(t)|}^2} + {{| \bar{\Omega}_{s}(t) |}^2}}$, ${\Xi _1}(t)=\sqrt{\varepsilon_{+}^{2}(t)+\Xi_{0}^{2}(t)}$, and
${\Xi _2}(t)=\sqrt{\varepsilon_{-}^{2}(t)+\Xi_{0}^{2}(t)}$.
The eigenvalues of $\bar{H_{0}}(t)$ corresponding to the eigenstates are,
\begin{eqnarray}
{E_0}(t) = 0,\quad {E_ \pm }(t) = \frac{\hbar }{2}{\varepsilon _ \pm }(t),
\label{zsq1}
\end{eqnarray}
with
\begin{eqnarray}
{\varepsilon _ \pm }(t) &=& ({\Delta} - {{i\Gamma } \mathord{/
 {\vphantom {{i\Gamma } 2}}
 \kern-\nulldelimiterspace} 2}) \pm \sqrt {{{({\Delta} - {{i\Gamma } \mathord{/
 {\vphantom {{i\Gamma } 2}}
 \kern-\nulldelimiterspace} 2})}^2} + \Xi _0^2(t)} .
 \label{zsq2}
\end{eqnarray}
In the STIRAP, the $| {{{ E}_0}(t)} \rangle$ eigenstate is a dark state without population on state $| 2 \rangle$. The protocol allows us to produce an effcient transfer from $| 1 \rangle$ to $| 3 \rangle$ following the
evolution of the dark state $| {{{ E}_0}(t)} \rangle$. Since $\bar H_0(t)$ is non-Hermitian, the instantaneous eigenstates of the Hamiltonian $\bar{H}^{\dagger}_{0}(t)$ have to be considered, which reads
\begin{eqnarray}
\begin{aligned}
| {{{\hat E}_0}(t)} \rangle  &= \frac{1}{{{\Xi _0}(t)}}[\bar{\Omega}_{s}(t)| 1 \rangle  - \bar{\Omega}^{\ast}_{p}(t)| 3 \rangle ],& \\
| {{{\hat E}_ + }(t)} \rangle  &= \frac{1}{{{\Xi _1}(t)}}[\bar{\Omega}_{p}(t)| 1 \rangle  + {{\hat \varepsilon }_ + }(t)| 2 \rangle
+\bar{\Omega}^{\ast}_{s}(t)| 3 \rangle ],&\\
| {{{\hat E}_ - }(t)} \rangle  &= \frac{1}{{{\Xi
_2}(t)}}[\bar{\Omega}_{p}(t)| 1 \rangle  + {{\hat \varepsilon }_ -
}(t)| 2 \rangle + \bar{\Omega}^{\ast}_{s}(t)| 3 \rangle ],&
\end{aligned}
\label{zsq3}
\end{eqnarray}
with ${{\hat \varepsilon }_ \pm }(t) = ({\Delta} + {{i\Gamma } \mathord{/ {\vphantom {{i\Gamma } 2}} \kern-\nulldelimiterspace} 2}) \pm \sqrt {{{({\Delta} + {{i\Gamma } \mathord{/ {\vphantom
{{i\Gamma } 2}} \kern-\nulldelimiterspace} 2})}^2} + \Xi_0^2(t)}$.

In order to realize the shortcuts to adiabaticity beyond RWA, we first write out three orthogonal projection operators corresponding to the relevant instantaneous eigenstates:
\begin{eqnarray}
\begin{aligned}
{\Pi _0}(t) &= | {{E_0}(t)} \rangle \langle {{{\hat E}_0}(t)} | = \frac{1}{{\Xi _0^2(t)}}\left [{\begin{array}{*{20}{c}}
{{{| \bar{\Omega}_{s}(t)|}^2}}&0& -\bar{\Omega}_{p}(t) \bar{\Omega}_{s}(t)\\
0&0&0\\
 - \bar{\Omega}^{\ast}_{p}(t)\bar{\Omega}^{\ast}_{s}(t)&0&{{{| \bar{\Omega}_{p}(t)|}^2}}
\end{array}}\right],&\\
{\Pi _1}(t) &= | {{E_1}(t)} \rangle \langle {{{\hat E}_1}(t)} | = \frac{1}{{\Xi _1^2(t)}}\left [{\begin{array}{*{20}{c}}
{{{| \bar{\Omega}_{p}(t) |}^2}}&\bar{\Omega}_{p}(t){\varepsilon _ + }(t)&\bar{\Omega}_{p}(t)\bar{\Omega}_{s}(t)\\
\bar{\Omega}^{\ast}_{p}(t){\varepsilon _ + }(t)&{\varepsilon _ + ^2(t)}&\bar{\Omega}_{s}(t){\varepsilon _ + }(t)\\
\bar{\Omega}^{\ast}_{p}(t)\bar{\Omega}^{\ast}_{s}(t)&\bar{\Omega}_{s}(t) {\varepsilon _ + }(t)&{{{| \bar{\Omega}_{s}(t)|}^2}}
\end{array}}\right],& \\
{\Pi _2}(t) &= | {{E_2}(t)} \rangle \langle {{{\hat E}_2}(t)} | = \frac{1}{{\Xi _2^2(t)}}\left [ {\begin{array}{*{20}{c}}
{{{| \bar{\Omega}_{p}(t) |}^2}}&\bar{\Omega}_{p}(t){\varepsilon _ - }(t)&\bar{\Omega}_{p}(t)\bar{\Omega}_{s}(t)\\
\bar{\Omega}^{\ast}_{p}(t){\varepsilon _ - }(t)&{\varepsilon _ - ^2(t)}&\bar{\Omega}_{s}(t){\varepsilon _ - }(t)\\
\bar{\Omega}^{\ast}_{p}(t)\bar{\Omega}^{\ast}_{s}(t)&\bar{\Omega}_{s}(t) {\varepsilon _ - }(t)&{{{| \bar{\Omega}_{s}(t) |}^2}}
\end{array}}\right].&
\end{aligned}
\label{zsq4}
\end{eqnarray}
By the same procedure, the Hermitian $\bar{H}_{CD}$ Hamiltonian becomes
\begin{small}
\begin{eqnarray}
\begin{aligned}
\bar{H}_{CD}(t) &=i\hbar \sum\limits_{j \ne k} {\sum {\frac{{{\Pi _j}(t){\partial _t} \bar{H}_{0}(t) {\Pi _k}(t)}}{{{E_k}(t) - {E_j}(t)}}} }& \\
&= \frac{{i\hbar }}{{\Xi _1^2(t)\Xi _2^2(t)}}\left [ {\begin{array}{*{20}{c}}
{{{| \bar{\Omega}_{p}(t) |}^2}B(t) + C(t)D(t)}&\bar{\Omega}_{p}(t)A(t) - 2i\Gamma \bar{\Omega}_{s}(t)G(t)&\bar{\Omega}_{p}(t)\bar{\Omega}_{s}(t)B(t) + C(t)F(t)\\
- \bar{\Omega}^{\ast}_{p}(t){A^*}(t) - 2i\Gamma \bar{\Omega}^{\ast}_{s}(t){G^*}(t)& - \Xi _0^2(t)B(t)& -\bar{\Omega}_{s}(t){A^*}(t) - 2i\Gamma \bar{\Omega}_{p}(t){G^*}(t)\\
- \bar{\Omega}^{\ast}_{p}(t)\bar{\Omega}^{\ast}_{s}(t){B^*}(t) - C(t){F^*}(t)&\bar{\Omega}^{\ast}_{s}(t)A(t) + 2i\Gamma \bar{\Omega}^{\ast}_{p}(t)G(t)&{{{| \bar{\Omega}_{s}(t)|}^2}B(t) - C(t)D(t)}
\end{array}}\right ],&
\end{aligned}
\label{zsq5}
\end{eqnarray}
\end{small}
with
\begin{small}
\begin{eqnarray}
\begin{array}{l}
A(t) = (2\Delta  - i\Gamma )\left [\bar{\Omega}^{\ast}_{p}(t)\dot{\bar{\Omega}}_{p}(t) + \bar{\Omega}_{s}(t)\dot{\bar{\Omega}}^{\ast}_{s}(t)\right ],\;\quad \\
B(t) = \bar{\Omega}^{\ast}_{p}(t)\dot{\bar{\Omega}}^{\ast}_{p}(t) + \bar{\Omega}_{s}(t)\dot{\bar{\Omega}}^{\ast}_{s}(t) - H.C.,\\
C(t) = \left [{| {{\varepsilon _ + }(t)} |^2} +{| {{\varepsilon _ - }(t)} |^2}+ 2\Xi _0^2(t) \right ] / \Xi _0^2(t),\quad \quad \\
D(t) = {| \bar{\Omega}_{p}(t) |^2} \left[ \bar{\Omega}^{\ast}_{s}(t)\dot{\bar{\Omega}}^{\ast}_{s}(t) - H.C. \right] + {| \bar{\Omega}_{s}(t) |^2}
\left [(\bar{\Omega}^{\ast}_{p}(t)\dot{\bar{\Omega}}_{p}(t) - H.C.\right ],\\
F(t)= \bar{\Omega}^{2}_{p}(t)\left [\bar{\Omega}_{s}(t)\dot{\bar{\Omega}}^{\ast}_{p}(t) - \bar{\Omega}^{\ast}_{p}(t)\dot{\bar{\Omega}}^{\ast}_{s}(t)\right]
+ \bar{\Omega}^{2}_{s}(t)\left [\bar{\Omega}^{\ast}_{s}(t)\dot{\bar{\Omega}}_{p}(t) - \bar{\Omega}_{p}(t)\dot{\bar{\Omega}}^{\ast}_{s}(t)\right],\quad \\
G(t)= \bar{\Omega}^{\ast}_{s}(t)\dot{\bar{\Omega}}_{p}(t) - \bar{\Omega}_{p}(t)\dot{\bar{\Omega}}^{\ast}_{s}(t).
\end{array}
\label{zsq6}
\end{eqnarray}
\end{small}
%%%%%%%%%%%%%%%%%%%%%%%%%%%%%%%%%%%%%%%%%%%%%%%%%%%%%%%%%
%figure 4
\begin{figure*}[tbp]
\centering
\includegraphics[angle=0,width=0.60\textwidth]{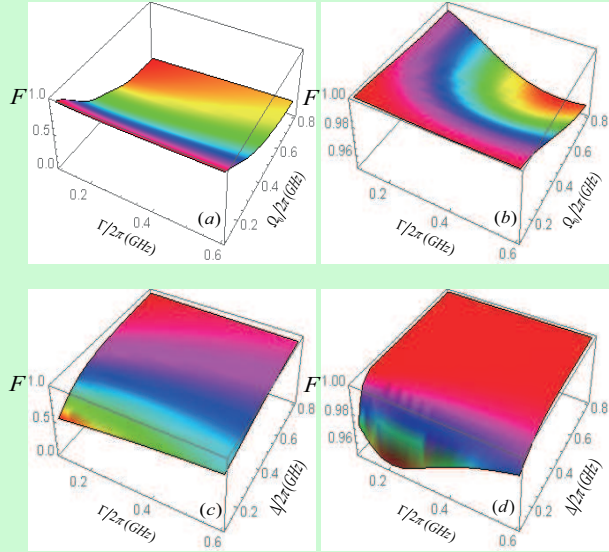}
\caption{Fidelity as a function of $\Gamma$, $\Omega_{0}$ and $\Delta$. The fidelity is defined as the population on the state $|3\rangle$ with initial state $|1\rangle$.
The population transfer is realized by using the STA technique under RWA but with counter-rotating terms [Figs. 4(a) and 4(c)], and without RWA [Figs. 4(b) and 4(d)]. In Figs. 4(a)
and 4(b), the parameters chosen are, $t_{f}=30 ns$, $\Omega_{0}= 2\pi \times [100,800] MHz$, $\Gamma= 2\pi \times [100,600] MHz$, $\omega_{p}=100 MHz$, $\omega_{s}=80 MHz$. In Figs. 4(c) and 4(d),
the parameters are set as: $t_{f}=30 ns$, $\Delta= 2\pi \times [100,800] MHz$, $\Gamma= 2\pi \times [100,600] MHz$, $\omega_{p}=100 MHz$, $\omega_{s}=80 MHz$. We can clearly find
that the fidelity without RWA is robust against the parameter fluctuations.}
\label{fig4}
\end{figure*}
Correspondently, we give the auxiliary driving for the shortcuts to adiabaticity under RWA
\begin{eqnarray}
{H_{CD}}(t) = i \hbar \left[ {\begin{array}{*{20}{c}}
0&{A(t)}&{C(t)}\\
{ - A(t)}&0&{ - B(t)}\\
{ - C(t)}&{B(t)}&0
\end{array}} \right],
\label{fulu}
\end{eqnarray}
where
\begin{eqnarray}
A(t) &=& \sin \theta (t) \dot \phi (t),\nonumber\\
B(t) &=& \cos \theta (t) \dot \phi (t),\label{zsq7}\\
C(t) &=& \dot \theta (t), \nonumber
\end{eqnarray}
with $\dot \phi (t)  = \frac{{{\dot \Omega' }}(t)(\Delta-i\Gamma / 2)-\Omega'(t)[\dot{\Delta}-i\dot{\Gamma}/2]}{2[\Omega'^{2}(t)+(\Delta_{p}-i\Gamma / 2)^{2}]}$,
 $ \dot \theta (t)  = \frac{{{{\dot \Omega_p}}(t){\Omega _s}(t) - {{\dot \Omega }_s}(t){\Omega_p}(t)}}{{\Omega^2(t)}}$, and $\Omega'(t)=\sqrt{\Omega_{p}^{2}+\Omega_{s}^{2}}$, respectively.

We would like to address that when the transitions  $| 1 \rangle \leftrightarrow | 2 \rangle$ and $| 2 \rangle \leftrightarrow | 3 \rangle$  are allowed, the transition $| 1 \rangle \leftrightarrow | 3 \rangle$
is in general forbidden. But for artificial atom ~\cite{artificialatom087001,artificialatom103604,artificialatom588,artificialatom193601,artificialatom840}, the forbiddenness of transition is lifted.

The Pump pulse and Stokes pulse in our numerical analysis can be described by ${\Omega _p}(t) = \Omega_{0} \exp[-(\frac{t-\tau-{{{t_f}} \mathord{/ {\vphantom {{{t_f}} 2}} \kern-\nulldelimiterspace} 2}}{T})^{2}]$
and ${\Omega _s}(t) = \Omega_{0} \exp[-(\frac{t+\tau-{{{t_f}} \mathord{/ {\vphantom {{{t_f}} 2}} \kern-\nulldelimiterspace} 2}}{T})^{2}]$. The Gaussion pulses have the same shapes and strengths but separated
by a delay of $2 \tau$. The entire interaction time of the system evolution is $t_{f}$. The constant loss rate $\Gamma $ is a constant. The initial conditions for the populations are ${P_1}(0) = 1$, ${P_2}(0) = 0$
and ${P_3}(0) = 0$. In Fig. 3(a) and 3(b),we can find that the shortcuts to adiabaticity technique under the RWA  does not works well if considering the counter-rotating effects, but beyond the RWA is not.

We use the fidelity between the finial state $ | {\Psi ( {{t_f}} )} \rangle $ and the target state $| 3 \rangle $ to characterize the population transfer efficiency (see Fig. \ref{fig4}), i.e.,
$F = {| {\langle {3} \mathrel{ | {\vphantom {3 {\Psi ( {{t_f}} )}}} \kern-\nulldelimiterspace} {{\Psi ( {{t_f}} )}} \rangle } |^2}$.
$F = {| {\langle {3} \mathrel{ | {\vphantom {3 {\Psi ( {{t_f}})}}} \kern-\nulldelimiterspace} {{\Psi ( {{t_f}} )}} \rangle }|^2}$. Figs. 4(a) and 4(c) show the fidelity beyond RWA, where we
use $H_{CD}(t)$ (Eq.~(\ref{fulu})) as the countdiabatic Hamiltonian. Figs. 4(b) and 4(d) are also for the results beyond RWA, but $\bar H_{CD}$  (Eq.~(\ref{zsq5})) is chosen as the
counterdiabatic Hamiltonian. The parameters used  in the numerical simulation are chosen as follows: {In Figs. 4(a) and 4(b), we choose $\Omega_{0}= 2 \pi \times[100,800] MHz$ and $\Gamma=2 \pi
\times[100,600] MHz$, respectively. In Figs. 4(c) and 4(d), the parameters are $\Delta= 2 \pi \times[100,800] MHz$ and $\Gamma=2 \pi \times[100,600] MHz$, respectively.
The final evolution time is $30ns$.} The numerical results show that the shortcuts to adiabaticity technique with the counterdiabatic Hamiltonian $H_{CD}(t)$ does not works well if the counter-rotating
terms are taken into account (see Figs. 4(a) and 4(c)). The fidelity is sensitive to the decay rate and pulse parameters. The required fidelities can be reached only in narrow range of parameters. This is due to the matching-lost between the adiabatic reference Hamiltonian $\bar H_0(t)$ and the counterdiabatic Hamiltonian $H_{CD}(t)$. To better the performance, we take $\bar H_{CD}(t)$ (Eq.~(\ref{zsq5})) as the counterdiabatic Hamiltonian to replace $H_{CD}(t)$, which are illustrated in Figs. 4(b) and 4(d). When the reference Hamiltonian and the counterdiabatic Hamiltonian are matching, the fidelity is not affected by any parameters in the coherent control field, and it is robust against the decay of the excited state.

It turned out that the previous shortcuts to adiabaticity technique under RWA does not work well if the counter-rotating terms are included. Therefore, the study of shortcuts to adiabaticity
technique beyond RWA is necessary. The robustness to the population loss of the excited state was investigated and it was shown that high fidelity can be achieved even with large decay rates. The decay of the excited
state has little effect on the population transfer. This is due to the fact that in the STIRAP the dark states are used, the excited state is almost not population during the evolution of the system.
We hope that this work may open a new door towards to the experimental realization of population transfer with decaying quantum system in the near future.

\subsection{Effective two-level system for large single-photon detuning }
%%%%%%%%%%%%%%%%%%%%%%%%%%%%%%%%%%%%%%%%%%%%%%%%%%%%%%%%%%%%%%%%%%%%%%%%%%%%%%%%%%%%%%%%%%%%%%%%%%%%%%%%%%%%%%%%%%%%%%%

When the detuning $\Delta (t) \gg \Omega_{p,s}(t)$, the middle state $| 2 \rangle$ can be eliminated adiabatically by setting $\dot{c}_{2} \simeq 0$ in Eq.~(\ref{lhx}). The
three-level system can be then reduced to an effective two-level system (with levels $| 1 \rangle$ and $| 3 \rangle$) ~\cite{NVV648,NVV013417,NVV763}:
\begin{eqnarray}
H_{eff}(t) = \frac{\hbar}{2} \left [ {\begin{array}{*{20}{c}}
-\Delta_{eff} (t)&{\Omega_{eff} (t)}\\
{\Omega^{\ast}_{eff} (t)}&{\Delta_{eff} (t) }
\end{array}} \right],
\label{eff11}
\end{eqnarray}
where $\Delta_{eff} (t)$ and $\Omega_{eff} (t)$ are effective detuning and Rabi frequency, respectively,
\begin{eqnarray}
\begin{aligned}
&\Delta_{eff} (t) = -\frac{ \bar{\Omega}_{p}(t) \bar{\Omega}_{p}(t)}{2\Delta-i\Gamma},&\\
&\Omega_{eff} (t) = \frac{|\bar{\Omega}_{p}(t)|^{2}-|\bar{\Omega}_{s}(t)|^{2}}{4\Delta-2i\Gamma}.&
\end{aligned}
\label{eff2}
\end{eqnarray}
In the large detuning limit, the three-level system reduces to an effective two-level system described by the Hamiltonian~(\ref{eff11}).
Once the effective two-level Hamiltonian is obtained, we can calculate the counter-adiabatic driving according to~\cite{LFWei023405}
\begin{eqnarray}
\begin{aligned}
H_{CD}(t) &= \frac{{i\hbar }}{{2M(t)}}\left [{\begin{array}{*{20}{c}}
{0}&{P(t)}\\
{{-P^{\ast}(t)}}&{{Q(t)}}
\end{array}}\right ],&
\end{aligned}
\label{eff3}
\end{eqnarray}
with
\begin{equation}
\begin{aligned}
P(t) &= \dot{\Omega}_{eff}(t) \Delta_{eff} (t)-\dot{\Delta}_{eff} (t) \Omega_{eff} (t),& \\
Q(t) &= \dot{\Omega}^{\ast}_{eff}(t)\Omega_{eff}(t)-\dot{\Omega}_{eff}(t)\Omega^{\ast}_{eff}, &\\
M(t) &= | \Omega_{eff} |^2+\Delta^{2}_{eff}.&
\label{lhw}
\end{aligned}
\end{equation}
Thus the total Hamiltonian reads $H(t)=H_{eff}(t)+H_{CD}(t)$. We can directly apply the effective two-level model in the analysis of our three-level system with large single-photon detuning. The
Pump pulse and Stokes pulse in our numerical analysis still can be chosen as ${\Omega _p}(t) = \Omega_{0} \exp[-(\frac{t-\tau-{{{t_f}} \mathord{/ {\vphantom {{{t_f}} 2}} \kern-\nulldelimiterspace} 2}}{T})^{2}]$
and ${\Omega _s}(t) = \Omega_{0} \exp[-(\frac{t+\tau-{{{t_f}} \mathord{/ {\vphantom {{{t_f}} 2}} \kern-\nulldelimiterspace} 2}}{T})^{2}]$. We now compare the performance of the above effective two-level system
shortcuts to adiabaticity and three-level system shortcuts to adiabaticity protocols. We set the detuning $\Delta=2 \pi \times 2.5 GHz$. With $\Omega_{0}=2 \pi \times 0.16 GHz$, $t_{f}=30 ns$, $T={t_f}/6$,
and $\tau={t_f}/10$, the dynamics described by the original Hamiltonian (\ref{lhy}) is not adiabatic at all, and population can not be completely transferred from $| 1 \rangle$ to $| 3 \rangle$.
Figs. 5(a) and 5(b) show the population transfer by shortcuts to adiabaticity of the effective two-level system and that of the three-level system, respectively. We can find that they are almost the same.
%figure 5
\begin{figure*}[tbp]
\centering
\includegraphics[angle=0,width=0.45\textwidth]{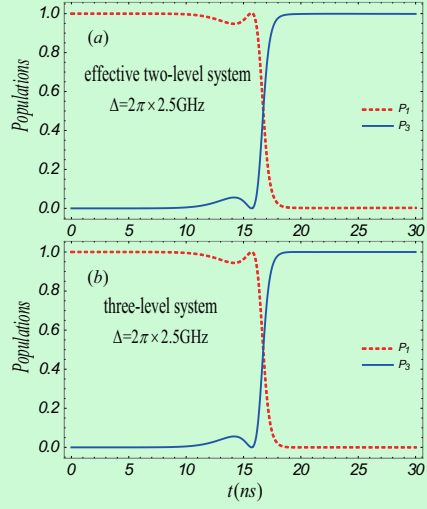}
\caption{The population evolution of the effective two-level system by shortcuts to adiabaticity (a), and the three-level system by shortcuts to adiabaticity (b). Here the population of level
$| 1 \rangle$ (dashed red), and $| 3 \rangle$ (solid blue) is presented. We set the parameters as: $t_{f}=30 ns$, $T={t_f}/6$, $\tau={t_f}/10$, $\Omega_{0}=2 \pi \times 0.16 GHz$,
$\Delta=2 \pi \times 2.5 GHz$, $\Gamma=2 \pi \times 0.16 GHz$, $\omega_{p}=0.1GHz$, $\omega_{s}=0.08GHz$. We can find that the population evolution given by the effective two-level model
and by the three-level model is almost the same.}
\label{fig5}
\end{figure*}

\section{Conclusion} \label{four}

Employing transitionless quantum driving, we have presented a method to design a shortcut to adiabaticity for a non-Hermitian system ~\cite{decay095303,decay053406} without RWA. We find that
the previous shortcuts to adiabaticity technique with RWA does not work well if the counter-rotating terms can not be neglected, instead the present scheme leads to a high performance to finish
the shortcut. Meanwhile we found that the decay of the excited state has little effect on the adiabatic population transfer in the three-level system. Our method is robust against the fluctuation
of parameters, as the fidelity is still unity for a wide range of $ \Gamma $, $\Omega_{0}$ and $\Delta$. In the case of large detuning, we reduce the three-level system to an effective two-level system by
using adiabatic elimination. The shortcuts to adiabaticity technique of effective two-level system is also studied and discussed.
\\
\\
\textbf{Funding}
\\
\\
This work is supported by National Natural Science Foundation of China (NSFC) under Grants Nos.~11534002, 61475033, 11775048, and 11705025, China Postdoctoral Science Foundation under Grant
Nos.~2016M600223 and 2017T100192, and the Fundamental Research Funds for the Central Universities under No.~2412017QD005.

\begin{thebibliography}{99}

\bibitem{Vitanov4535} N. V. Vitanov, K. A. Suominen, and B. W. Shore, ``Creation of coherent atomic superpositions by fractional stimulated Raman adiabatic passage," J. Phys. B \textbf{32}, 4535 (1999).

\bibitem{shiyan023821} Y. X. Du, Z. T. Liang, W. Huang, H. Yan, and S. L. Zhu, ``Experimental observation of double coherent stimulated Raman adiabatic passages in three-level $\Lambda$ systems in a cold atomic ensemble,"
Phys. Rev. A \textbf{90}, 023821 (2014).

\bibitem{Wei113601} L. F. Wei, J. R. Johansson, L. X. Cen, S. Ashhab, and F. Nori, ``Controllable coherent population transfers in superconducting qubits for quantum computing," Phys. Rev. Lett. \textbf{100}, 113601 (2008).

\bibitem{Halfmann7068} T. Halfmann and K. Bergmann, ``Coherent Population Transfer and Dark Resonances in $SO_{2}$," J. Chem. Phys. \textbf{104}, 7068 (1996).

\bibitem{Bergmann1003} K. Bergmann, H. Theuer, and B. W. Shore, ``Coherent Population Transfer Among Quantum States of Atoms and Molecules," Rev. Mod. Phys. \textbf{70}, 1003 (1998).


%adiabatic
\bibitem{LG033419} L. Giannelli and E. Arimondo, ``Three-level superadiabatic quantum driving," Phys. Rev. A \textbf{89}, 033419 (2014).

\bibitem{Sun033408} Y. Sun and H. Metcalf, ``Nonadiabaticity in stimulated Raman adiabatic passage," Phys. Rev. A \textbf{90}, 033408 (2014).

\bibitem{Wiebe013024} N. Wiebe and N. S. Babcock, ``Improved error-scaling for adiabatic quantum evolutions," New J. Phys. \textbf{14}, 013024 (2012).

\bibitem{Takahashi315304} K. Takahashi, ``How fast and robust is the quantum adiabatic passage?" J. Phys. A: Math. Theor. \textbf{46}, 315304 (2013).

\bibitem{Torosov043418} B. T. Torosov and N. V. Vitanov, ``Composite Stimulated Raman Adiabatic Passage," Phys. Rev. A \textbf{87}, 043418 (2013).

%STA
\bibitem{chenxi033405} X. Chen and J. G. Muga, ``Engineering of fast population transfer in three-level systems," Phys. Rev. A \textbf{86}, 033405 (2012).

\bibitem{STA123003} X. Chen, I. Lizuain, A. Ruschhaupt, D. Gu\'{e}ry-Odelin, and J. G. Muga, ``Shortcut to Adiabatic Passage in Two- and Three-Level Atoms," Phys. Rev. Lett. \textbf{105}, 123003 (2010).

\bibitem{STA9937} M. Demirplak and S. A. Rice, ``Adiabatic Population Transfer with Control Fields," J. Phys. Chem. A \textbf{107}, 9937 (2003).

\bibitem{STA25707} X. B. Huang, Y. H. Chen, and Z. Wang, ``Fast generation of three-qubit Greenberger-Horne-Zeilinger state based on the Lewis-Riesenfeld invariants in coupled cavities," Sci. Rep. \textbf{6}, 25707 (2016).

\bibitem{STA1630} J. Sun, S. F. Lu, and F. Liu, ``Speedup in adiabatic evolution based quantum algorithms," Sci. China-Phys. Mech. Astron. \textbf{55}, 1630 (2012).

\bibitem{chen043402} S. Ib\'{a}$\tilde{n}$ez, X. Chen, and J. G. Muga, ``Improving shortcuts to adiabaticity by iterative interaction pictures," Phys. Rev. A \textbf{87}, 043402 (2013).

\bibitem{Paul052303} K. Paul and A. K. Sarma, ``High-fidelity entangled Bell states via shortcuts to adiabaticity," Phys. Rev. A \textbf{94}, 052303 (2016).

\bibitem{xiayan052109} Y. H. Chen, Y. Xia, Q. C. Wu, B. H. Huang, and J. Song, ``Method for constructing shortcuts to adiabaticity by a substitute of counterdiabatic driving terms," Phys. Rev. A \textbf{93}, 052109 (2016).

\bibitem{Campbell177206} S. Campbell, G. De Chiara, M. Paternostro, G. M. Palma, and R. Fazio, ``Shortcut to Adiabaticity in the Lipkin-Meshkov-Glick Model," Phys. Rev. Lett. \textbf{114}, 177206 (2015).

\bibitem{Okuyama070401} M. Okuyama and K. Takahashi, ``From Classical Nonlinear Integrable Systems to Quantum Shortcuts to Adiabaticity," Phys. Rev. Lett. \textbf{117}, 070401 (2016).

\bibitem{STA052303} S. He, S. L. Su, D. Y. Wang, W. M. Sun, C. H. Bai, A. D. Zhu, H. F. Wang, and S. Zhang, ``Efcient
shortcuts to adiabatic passage for three-dimensional entanglement generation via transitionless quantum driving," Sci. Rep. \textbf{6}, 30929 (2016).

\bibitem{Kumar10628} K. S. Kumar, A. Veps\"{a}l\"{a}inen, S. Danilin, and G. S. Paraoanu, ``Stimulated Raman adiabatic passage in a three-level superconducting circuit," Nat. Commun. \textbf{7}, 10628 (2016).

%TQD

\bibitem{TQD053408} S. Martnez-Garaot, E. Torrontegui, X. Chen, and J. G. Muga,  ``Shortcuts to adiabaticity in three-level systems using Lie transforms," Phys. Rev. A \textbf{89}, 053408 (2014).

\bibitem{TQD015025} T. Opatrn$\acute{y}$ and K. M{\o}mer, ``Partial suppression of nonadiabatic transitions," New J. Phys. \textbf{16}, 015025 (2014).

\bibitem{TQD100403} S. Ib\'{a}$\tilde{n}$ez, X. Chen, E. Torrontegui, J. G. Muga, and  A. Ruschhaupt, ``Multiple Schr\"{o}dinger Pictures and Dynamics in Shortcuts to Adiabaticity," Phys. Rev. Lett. \textbf{109}, 100403 (2012).

\bibitem{TQD052109} Y. H. Chen, Q. C. Wu, B. H. Huang, Y. Xia, and J. Song,  ``Method for constructing shortcuts to adiabaticity by a substitute of counterdiabatic driving terms," Phys. Rev. A \textbf{93}, 052109 (2016).

\bibitem{Takahashi062117} K. Takahashi, ``Transitionless quantum driving for spin systems," Phys. Rev. E \textbf{87}, 062117 (2013).

\bibitem{LFWei015204} X. Shi and L. F. Wei, ``High-efficiency single-photon Fock state production by transitionless quantum driving, Laser," Phys. Lett. \textbf{12}, 015204 (2015).

\bibitem{LRchenxi062116} X. Chen, E. Torrontegui, and J. G. Muga, ``Lewis-Riesenfeld invariants and transitionless quantum driving," Phys. Rev. A \textbf{83}, 062116 (2011).

\bibitem{Songxueke052324} X. K. Song, Q. Ai, J. Qiu, and F. G. Deng, ``Physically feasible three-level transitionless quantum driving with multiple Schr\"{o}dinger dynamics," Phys. Rev. A \textbf{93}, 052324 (2016); X. K. Song, H. Zhang, Q. Ai, J. Qiu, and F. G. Deng, ``Shortcuts to adiabatic holonomic quantum computation in decoherence-free subspace with transitionless quantum driving algorithm," New J. Phys. \textbf{18}, 023001 (2016).

\bibitem{Berry2009} M. V. Berry, ``Transitionless quantum driving," J. Rhys. A: Math. Theor \textbf{42}, 365303 (2009).

\bibitem{Masuda043434} S. Masuda and K. Nakamura, ``Acceleration of adiabatic quantum dynamics in electromagnetic fields," Phys. Rev. A \textbf{84}, 043434 (2011).

\bibitem{Kiely115501} A. Kiely and A. Ruschhaupt, ``Inhibiting unwanted transitions in population transfer in two- and three-level quantum systems," J. Phys. B \textbf{47}, 115501 (2014).

\bibitem{Masuda033621} S. Masuda and S. A. Rice, ``Rapid Coherent Control of Population Transfer in Lattice Systems," Phys. Rev. A \textbf{89}, 033621 (2014).

%Non-Hermitian

\bibitem{NHMoiseyev2011} N. Moiseyev, \textit{Non-Hermitian Quantum Mechanics} (Cambridge University, Cambridge, 2011).

\bibitem{PTBender957} C. M. Bender and S. Boettcher, ``Real Spectra in Non-Hermitian Hamiltonians Having P-T Symmetry," Phys. Rev. Lett. \textbf{80},5243(1998);
C. M. Bender, ``Making sense of non-Hermitian Hamiltonians," Rep. Prog. Phys. \textbf{70}, 957 (2007).

\bibitem{PTMostafazadeh7081} A. Mostafazadeh, ``Pseudo-Hermiticity versus PT symmetry: The necessary condition for the reality of the spectrum of a non-Hermitian Hamiltonian," J. Math. Phys. \textbf{43}, 205 (2002);
A. Mostafazadeh, ``Exact PT-symmetry is equivalent to Hermiticity," J. Phys. A \textbf{36},7081 (2003).

\bibitem{Uzdin415304} R. Uzdin, U. G\"{u}nther, S. Rahav, and N. Moiseyev, ``Time-dependent Hamiltonians with 100\% evolution speed efficiency," J. Phys. A: Math. Theor. \textbf{45}, 415304 (2012).

\bibitem{Hang083604} C. Hang, G. Huang, and V. V. Konotop, ``PT-Symmetry with a System of Three-Level Atoms," Phys. Rev. Lett. \textbf{110}, 083604 (2013).

\bibitem{Ibanez033403} S. Ib\'{a}\~{n}ez and J. G. Muga, ``Adiabaticity condition for non-Hermitian Hamiltonians," Phys. Rev. A \textbf{89}, 033403 (2014).

\bibitem{Reyes444027} S. A. Reyes, F. A. Olivares, and L. Morales-Molina, ``Landau-Zener-St\"{u}ckelberg interferometry in PT-symmetric optical waveguides," J. Phys. A: Math. Theor. \textbf{45}, 444027 (2012).

%RWA

\bibitem{Chan065507} S. Chan, M. D. Reid, and Z. Ficek, ``Entanglement evolution of two remote and non-identical Jaynes-Cummings atoms," J. Phys. B \textbf{42}, 065507(2009).

\bibitem{Irish259901} E. K. Irish, ``Generalized rotating-wave approximation for arbitrarily large coupling," Phys. Rev. Lett. \textbf{99}, 173601 (2007).

\bibitem{Klimov063811} A. B. Klimov, I. Sainz, and S. M. Chumakov, ``Resonance expansion versus the rotating-wave approximation," Phys. Rev. A \textbf{68}, 063811 (2003).

\bibitem{Liu54003} T. Liu, K. L. Wang, and M. Feng, ``The generalized analytical approximation to the solution of the single-mode spin-boson model without rotating-wave approximation," Europhys. Lett. \textbf{86}, 54003 (2009).

\bibitem{Ashhab042311} S. Ashhab and F. Nori, ``Qubit-oscillator systems in the ultrastrong-coupling regime and their potential for preparing nonclassical states," Phys. Rev. A \textbf{81}, 042311 (2010).

\bibitem{Casanova263603} J. Casanova, G. Romero, I. Lizuain, J. J. Garc\'{\i}a-Ripoll, and E. Solano, ``Deep Strong Coupling Regime of the Jaynes-Cummings Model," Phys. Rev. Lett. \textbf{105}, 263603 (2010).

\bibitem{Fedorov060503} A. Fedorov, A. K. Feofanov, P. Macha, P. Forn-D\'{\i}az, C. J. P. M. Harmans, and J. E. Mooij, ``Strong Coupling of a Quantum Oscillator to a Flux Qubit at Its Symmetry Point," Phys. Rev. Lett. \textbf{105}, 060503 (2010).

\bibitem{Irish173601} E. K. Irish, ``Generalized Rotating-Wave Approximation for Arbitrarily Large Coupling," Phys. Rev. Lett. \textbf{99}, 173601 (2007).

\bibitem{Bourassa032109} J. Bourassa, J. M. Gambetta, A. A. Abdumalikov, Jr., O. Astafiev, Y. Nakamura, and A. Blais, ``Ultrastrong coupling regime of cavity QED with phase-biased flux qubits," Phys. Rev. A \textbf{80}, 032109 (2009).

\bibitem{Larson033601} J. Larson, ``Absence of Vacuum Induced Berry Phases without the Rotating Wave Approximation in Cavity QED," Phys. Rev. Lett. \textbf{108}, 033601 (2012).

\bibitem{Sun012107} Z. Sun, J. Ma, X. Wang, and F. Nori, ``Photon-assistedLandau-Zenertransition:Role of coherent superposition states," Phys. Rev. A \textbf{86}, 012107 (2012).

\bibitem{You589} J. Q. You, and F. Nori, ``Atomic physics and quantum optics using superconducting circuits," Nature \textbf{474}, 589 (2011).

\bibitem{Niemczyk772} T. Niemczyk, F. Deppe, H. Huebl, E. P. Menzel, F. Hocke, M. J. Schwarz, J. J. Garcia-Ripoll, D. Zueco, T. H$\ddot{u}$mmer, E. Solano,  A. Marx, and R. Gross. ``Circuit quantum electrodynamics in the ultrastrong-coupling regime," Nature Phys. \textbf{6}, 772 (2010).

%decay
\bibitem{decay053406} K. Paul and A. K. Sarma, ``Shortcut to adiabatic passage in a waveguide coupler with a complex-hyperbolic-secant scheme," Phys. Rev. A \textbf{91}, 053406 (2015).

\bibitem{decay095303}M. B. Kenmoe, S. E. M. Tchouobiap, C. K. Sadem, A. B. Tchapda, and L. C. Fai, ``Non-adiabatic and adiabatic transitions at level crossing with decay: two- and three-level systems," J. Phys. A \textbf{48}, 095303 (2015).

\bibitem{decay13727} R. Schilling, M. Vogelsberger, and D. A. Garanin, ``Nonadiabatic transitions for a decaying two-level system: geometrical and dynamical contributions," J. Phys. A: Math. Gen. \textbf{39}, 13727 (2006).

\bibitem{xjdecay043001} R. Garcia Fernandez, A. Ekers, L. P. Yatsenko, N. V. Vitanov, and K. Bergmann, ``Control of Population Flow in Coherently Driven Quantum Ladders," Phys. Rev. Lett. \textbf{95}, 043001 (2005).

\bibitem{xjdecay053415} I. I. Boradjiev and N. V. Vitanov, ``Stimulated Raman adiabatic passage with unequal couplings: Beyond two-photon resonance," Phys. Rev. A \textbf{81}, 053415 (2010).

\bibitem{chenxi023415} S. Ib\'{a}nez, S. Martinez-Garaot, X. Chen, E. Torrontegui, and J. G. Muga, ``Shortcuts to adiabaticity for non-Hermitian systems," Phys. Rev. A \textbf{84}, 023415 (2011).

\bibitem{chenxi062163} S. Ib\'{a}$\tilde{n}$ez, Y. C. Li, X. Chen, and J. G. Muga, ``Pulse design without the rotating-wave approximation," Phys. Rev. A \textbf{92}, 062136 (2015).

\bibitem{xjdecay015006} N. V. Vitanov, A. A. Rangelov, B. W. Shore, and K. Bergmann, ``Stimulated Raman adiabatic passage in physics, chemistry, and beyond," Rev. Mod. Phys. \textbf{89}, 015006 (2017).

%artificialatom

\bibitem{artificialatom087001} Y. X. Liu, J. Q. You, L. F. Wei, C. P. Sun, and F. Nori, ``Optical Selection Rules and Phase-Dependent Adiabatic State Control in a Superconducting Quantum Circuit," Phys. Rev. Lett. \textbf{95},087001 (2005).

\bibitem{artificialatom103604} L. Zhou, L. P. Yang, Y. Li, and C. P. Sun, ``Quantum Routing of Single Photons with a Cyclic Three-Level System," Phys. Rev. Lett. \textbf{111}, 103604 (2013).

\bibitem{artificialatom588} O. Astafiev, K. Inomata, A. O. Niskanen, T. Yamamoto, Yu. A. Pashkin, Y. Nakamura, and J. S. Tsai, ``Single artificial-atom lasing," Nature \textbf{449}, 588(2007).

\bibitem{artificialatom193601} A. A. Abdumalikov, Jr., O. Astafiev, A. M. Zagoskin, Yu. A. Pashkin, Y. Nakamura, and J. S. Tsai, ``Electromagnetically Induced Transparency on a Single Artificial Atom," Phys. Rev. Lett. \textbf{104}, 193601(2010).

\bibitem{artificialatom840} O. Astafiev, A. M. Zagoskin, A. A. Abdumalikov, Jr., Yu. A. Pashkin, T. Yamamoto, K. Inomata, Y. Nakamura, and J. S. Tsai, ``Resonance Fluorescence of a Single
Artificial Atom," Science \textbf{327}, 840 (2010).

%effective two-level

\bibitem{NVV648} N. V. Vitanov and S. Stenholm, ``Analytic properties and effective two-level problems in stimulated Raman adiabatic passage," Phys. Rev. A \textbf{55}, 648 (1997).

\bibitem{NVV013417} G. S. Vasilev, A. Kuhn, and N. V. Vitanov, ``Optimum pulse shapes for stimulated Raman adiabatic passage," Phys. Rev. A \textbf{80}, 013417 (2009).

\bibitem{NVV763} N. V. Vitanov, T. Halfmann, B. W. Shore, and K. Bergmann, ``Laser-induced population transfer by adiabatic passage techniques," Annu. Rev. Phys. Chem. \textbf{52}, 763 (2001).

\bibitem{LFWei023405} J. W. Chen and L. F. Wei, ``Implementation speed of deterministic population passages compared to that of Rabi pulses," Phys. Rev. A \textbf{91}, 023405 (2015).
\end{thebibliography}
\end{document}